\documentclass{article}
\usepackage[utf8]{inputenc}
\usepackage[round,authoryear]{natbib}
\usepackage{libertine}
\usepackage{graphicx}
\usepackage{epigraph}
\usepackage{comment}
\usepackage[table]{xcolor}
\usepackage{amsfonts,amsmath,amssymb,amsthm,url, mathrsfs, graphicx, pxfonts, xcolor}
\usepackage{hyperref}

\usepackage{physics}
 \usepackage[labelfont=bf]{caption}
\usepackage{multirow}

\newcommand{\x}[1]{{#1}}
\newcommand{\y}[1]{{#1}}
\newcommand{\ec}[1]{{#1}}

\newcommand{\xx}[1]{{#1}}

\newlength{\bibitemsep}
\newlength{\bibparskip}
\let\oldthebibliography\thebibliography
\renewcommand\thebibliography[1]{%
  \oldthebibliography{#1}%
  \setlength{\parskip}{\bibitemsep}%
  \setlength{\itemsep}{\bibparskip}%
}

\newcommand{\Hilbert}{\mathscr{H}}

\title{Decoherence, Branching, and the Born Rule \\ in a Mixed-State Everettian Multiverse\thanks{This work is fully collaborative; the authors are listed anti-alphabetically.}
}
\author{Eugene Y. S. Chua\thanks{School of Humanities, Nanyang Technological University Singapore. Website: www.eugenechua.com. Email: eugene.chuays@ntu.edu.sg} \;\thanks{Division of the Humanities and Social Sciences, California Institute of Technology.} \, and Eddy Keming Chen\thanks{Department of Philosophy, University of California, San Diego. Website: www.eddykemingchen.net. Email: eddykemingchen@ucsd.edu}}
\date{Accepted version, \textit{Synthese}}

\begin{document}

\maketitle

\begin{abstract}
  \noindent In Everettian quantum mechanics, justifications for the Born rule appeal to self-locating uncertainty or decision theory. Such justifications have focused exclusively on a pure-state Everettian multiverse, represented by a wavefunction. Recent works in quantum foundations suggest that it is viable to consider a mixed-state Everettian multiverse, represented by a (mixed-state) density matrix. Here, we discuss the conceptual foundations for decoherence and branching in a mixed-state multiverse, and extend arguments for the Born rule to this setting. This extended framework provides a unification of `classical' and `quantum' probabilities, and additional theoretical benefits, for the Everettian picture.
\end{abstract}

\begingroup
\tableofcontents
\endgroup

\section{Introduction}

Everettian quantum mechanics (EQM) is a minimalist interpretation of quantum mechanics with some counter-intuitive features (\cite{sep-qm-manyworlds, sep-qm-everett}). Instead of collapsing the quantum state or adding extra variables to obtain a definite outcome for each experiment, it proposes to take unitary quantum mechanics as fundamental and replace our single-world ontology with a multiverse, where every possible outcome of an experiment is realized in some branch (a parallel world). 

 EQM faces two main issues, one metaphysical and the other epistemological. The metaphysical issue concerns the ontology of EQM. How do we obtain the appearance of a classical world, with definite records and observers, from the quantum state? A much discussed solution appeals to decoherence, with its ability to suppress interference and give rise to an ``emergent multiverse'' (\cite{Wallace2012}). The universal quantum state evolves into one with many branches, each representing an emergent (quasi-)classical world. 

The epistemological issue concerns the understanding of probability in EQM. A key postulate of quantum mechanics, and a crucial element of its empirical confirmation, is the Born rule: the probability of observing a certain outcome is given by the squared amplitude of the quantum state. How should we  make sense of this probability when every measurement outcome occurs on some branch of the Everettian multiverse, and what justifies the interpretation of the squared amplitudes as probabilities? \xx{Several responses have been proposed.} The decision-theoretic program understands probability in terms of the betting preferences of agents within the multiverse, and uses decision-theoretic considerations to prove that the agent's credences must satisfy the Born rule, on pain of irrationality (e.g. \cite{deutsch1999}, \cite{Wallace2012}). The Sebens-Carroll (\citep{sebenscarroll2016}) and McQueen-Vaidman (\citep{mcqueenvaidman2018})  programs understand probability in terms of self-locating uncertainty of a localized agent on some branch, employing certain epistemic principles to prove that the agent's self-locating uncertainty must satisfy the Born rule. 

However, these defenses and justifications of EQM have an apparent limitation. They focus exclusively on the case of a universal pure state, where the quantum state of the multiverse is represented by a wavefunction. Defenders of EQM, like many other realist interpreters, regard the universal pure state as representing something objective and mind-independent. However, recent works in quantum foundations (\cite{durr2005role, maroney2005density, allori2013predictions, Wallace2012, chen2018, robertson2022search}) suggest that the above approach to realism, based on the wavefunction, is not the only possibility for realism about the quantum state. It's also viable -- and arguably more theoretically attractive in certain respects (\cite{chen2018, chen2018HU, chen2024DMR}) -- to take a realist stance based on the density matrix. On this view, we can associate (possibly mixed-state) density matrices, rather than (necessarily pure-state) wavefunctions, to isolated systems and even to the entire universe. While density matrices are conventionally used to represent ignorance about some underlying wavefunction, \textit{it's also possible to regard density matrices as physically fundamental}. On the new picture, the universe as a whole can be aptly represented by a fundamental density matrix evolving unitarily according to the von Neumann equation, in contrast to the standard picture of a wavefunction evolving unitarily according to the Schr\"odinger equation. If the fundamental density matrix in this new realist picture is mathematically equivalent to that of the ``ignorance'' density matrix in the standard picture, the two theories will be empirically equivalent, since they make the same statistical predictions for all experiments.  

All wavefunctions correspond to some pure-state density matrices, but not all density matrices have corresponding wavefunctions. Thus, realism based on the density matrix allows for more quantum states than realism based on the wavefunction. \x{Theories that allow fundamental mixed states and those that disallow them have different fundamental state spaces, rendering them physically distinct.}\footnote{\x{There are two ways to see they are physically distinct. First, according to the ontological models framework developed by \citet{harrigan2010einstein}, such differences result in different choices of $\Lambda$, the ontic space. Different ontic spaces lead to different quantum theories. Second, according to many current accounts of laws of nature, differences in which fundamental states are allowed can be construed as differences in the physical laws. See \citet{chenCUP} for an overview.} }  Following \citet{chen2018, chen2019realism}, we call this new picture \textit{Density Matrix Realism} (DMR) and the old one \textit{Wavefunction Realism} (WFR). We denote the Everettian versions of DMR and WFR as DMR\textsubscript{E} and WFR\textsubscript{E} respectively. (Note: this is a wider conception of quantum state realism than \citet{AlbertEQM} and \citet{ney2021world}.)

It is an open question whether standard arguments for branching and the Born rule generalize from WFR\textsubscript{E} to DMR\textsubscript{E}. If there's no generalization available,  WFR\textsubscript{E} might still be preferable to DMR\textsubscript{E}, since the former -- not the latter -- 
\xx{has detailed responses to the problems of ontology and probability, though these proposals remain subjects of ongoing debate and far from universally accepted.}\footnote{\xx{The proposed solutions to the problems of ontology and probability in WFR\textsubscript{E} remain highly contested. For example, debates continue about the ontological status of wave functions  (e.g. \cite{wallace2010quantum}, \cite[ch.7]{lewis2016quantum}, \cite{chen2019realism}, \cite{wallace2022stating, wallace2024real}, \cite{franklin2024incoherent}) and the nature of probability (e.g. 
\cite{greaves2004understanding}, \cite{Baker2007}, \cite{kent2010}, \cite{dawid2015many}, \cite{saunders2021, saunders2024finite}, \cite{adlam2024against}). Our contribution is not to resolve these debates but to extend the proposed solutions---insofar as they are successful---from the special case of pure states to the more general case of (possibly mixed-state) density matrices, while also highlighting the theoretical virtues of this generalization.}} Here, we argue that the standard decoherence-based justification for branching in WFR\textsubscript{E} can be extended to DMR\textsubscript{E} (\S2). Moreover, we show how the justifications for the Born rule in terms of self-location -- the Sebens-Carroll program (\S3.1) and the McQueen-Vaidman program (\S3.2) -- and the decision-theoretic program \x{(as exemplified by Deutsch's proof, \S3.3)} do not depend crucially on WFR\textsubscript{E}, but can also extend readily to DMR\textsubscript{E}. The ideas we develop here may be helpful to other such generalizations.

\x{We believe that this project is of interest to philosophers of science. First, several proposals in quantum cosmology (\citet{page1986density, page2008no}, \citet{barvinsky2006cosmological}) and quantum gravity (\citet{hawking1976breakdown, hawking1982unpredictability}) posit a universal mixed state. They are compatible with DMR but incompatible with WFR. Hence, it is of scientific interest whether we can generalize EQM to DMR.\footnote{\x{Here we focus on generalizing unitary versions of EQM. To accommodate, for instance, the pure-to-mixed transition in certain interpretations of black hole evaporation (\citet{hawking1976breakdown, hawking1982unpredictability}) one may take a further step and consider non-unitary versions, such as the proposal of \citet{cuffaro2023open}.}} 

Second, rivals to EQM, such as Bohmian mechanics and spontaneous collapse theories, have been successfully extended to DMR (\citet{durr2005role}, \citet{allori2013predictions}, \citet{chen2024DMR}). However, in EQM, the generalizations of solutions to the ontology and probability problems have been absent. If their extensions were impossible, EQM would be less generalizable than its rivals in this respect, an interesting result that bears on their overall comparisons. Our results show that EQM is equally generalizable to DMR.   

Third, Everettians themselves have considered whether EQM can be generalized to DMR (\citet{deutsch1999}, p.3136; \citet{Wallace2012}, \S10.5). Our project should be of interest to them. Specifically, we show that Everettian solutions to the problems of ontology and probability are \textit{robust}. We find that branching requires decoherence but decoherence does not require a universal pure state. The story of decoherence applies to both pure and mixed states, which has been underappreciated in the literature.  Moreover, the fundamental insights of the decision-theoretic and self-location justifications for the Born rule also apply to universal mixed states. 

Fourth, DMR is relevant to recent discussions in the foundations of thermodynamics, such as the proposals of \citet{McCoySMS}, \citet{chen2018}, and \citet{robertson2022search}. For example, DMR but not WFR is compatible with a new theoretical package -- the Wentaculus -- which provides a unified explanation for quantum phenomena and the thermodynamic arrow of time (\citet{chen2018, chen2018HU, chen2024detlef}).
The Wentaculus uses the Past Hypothesis to select a mixed-state density matrix of the universe, the normalized projection onto the Past-Hypothesis subspace. It is argued that this choice leads to significant theoretical benefits, such as the reduction of statistical probabilities to quantum probabilities (\citet{chen2018valia}), the elimination of nomic vagueness (\citet{chen2018NV}), and the realization of strong determinism (\citet{chen2022strong, ChenNature2023}).  Insofar as one accepts such arguments, one can now extend those benefits to EQM.  

Fifth, our results establish DMR\textsubscript{E} as a viable version of EQM and a viable rival to WFR\textsubscript{E}, by showing the former capable of tackling EQM's issues of ontology and probability, via some of the same resources as the latter. Now, DMR\textsubscript{E} and WFR\textsubscript{E} are physically different, since they are compatible with different sets of universal quantum states. This leads to a case of in-principle empirical underdetermination: \textit{by Everettians' own lights},  there is an open question which version of EQM is the correct one. That is an interesting fact in itself; it is also relevant to discussions of scientific realism in the quantum domain. 

Finally, DMR is a coherent framework to formulate solutions to the measurement problem, and DMR\textsubscript{E} is an intelligible version of unitary quantum theory. We believe that we should investigate each framework and theory seriously, so that we can be better informed about the space of possible physical theories and their inter-theoretic relations. 
}

\section{Decoherence and Branching}

\subsection{Decoherence at the Subsystem Level}

We start with a brief review of decoherence at the subsystem level for WFR\textsubscript{E}. Here we mostly follow \citet{schlosshauer2007decoherence}.

Consider a universal pure state $\Psi$ describing a system $S$ interacting with the environment $E$.  Given a system $S$ in a microscopic superposition of states $S_n$: 
\begin{equation}
    |S\rangle = \sum_{n} c_n |S_n\rangle
\end{equation}
interacting with $E$, at some time $t$ after the interaction, the universal state will become a macroscopic superposition: 
\begin{equation}\label{universal}
    |S+E\rangle = \sum_{n}c_n |S_n\rangle |E_n(t)\rangle
\end{equation}
\ec{where $|E_n(t)\rangle$ is the macroscopic `pointer state' associated with $S_n$.} For a simple case, consider $n = 2$. The reduced density matrix of subsystem $S$ is: 
    \begin{equation}\label{subsystem}
    \begin{split}
      \rho_S = Tr_E (\rho_{S+E}) & = Tr_E|S+E\rangle \langle S+E| \\
    & = |c_1|^2|S_1\rangle \langle S_1| + |c_2|^2|S_2\rangle \langle S_2| \\
    & + c_1c_2^*|S_1\rangle \langle S_2| \langle  E_2(t)|E_1(t)\rangle + c_1^*c_2 |S_2\rangle \langle S_1| \langle E_1(t)|E_2(t)\rangle  
    \end{split}
    \end{equation}
The last two terms (the off-diagonal ones) represent the interference between the two macroscopically superposed states, and depends partly on $\langle E_1(t)|E_2(t)\rangle$ and $\langle E_2(t)|E_1(t)\rangle$. \x{We refer to such terms as ``cross terms'' or ``c.t.'' for short.} Generically, $\langle E_i(t)|E_j(t)\rangle$ quantifies the difference between two states of the environment. Due to their large numbers of degrees of freedom, the states of the macroscopic environment become rapidly approximately orthogonal under Schrödinger evolution, such that
\begin{equation}
    \langle E_i(t)|E_j(t)\rangle \propto e^{-t/\tau_d}, i \neq j
\end{equation}
holds, where $\tau_d$ is the characteristic decoherence timescale to be empirically determined for specific systems. Over time, $\langle E_1(t)|E_2(t)\rangle$ and $\langle E_2(t)|E_1(t)\rangle$ approach zero, so that: 
\begin{equation}
    \rho_S \approx |c_1|^2|S_1\rangle \langle S_1| + |c_2|^2|S_2\rangle \langle S_2|
\end{equation}
In other words, any measurement on the system $S$, entangled with $E$, effectively ignores the quantum interference between the macroscopically superposed component states. 

\x{In this presentation, the cross terms show up only at the subsystem level (\ref{subsystem}). With decoherence, such terms rapidly become vanishingly small. However, we do not see the analogous effect at the universal level (\ref{universal}), because cross terms never appear in the universal pure state. Does decoherence apply to the universe as a whole? Yes. There is another way of thinking about decoherence, in terms of the decoherent-histories approach (DH), according to which we do not need to start with a reduced density matrix. On that approach, cross terms show up at both the universal level and the subsystem level. The DH criterion for decoherence does not presuppose a split of the universe into a system and its environment. We turn to DH next and show that it applies in the same way to DMR and WFR.}

\x{\subsection{Decoherent Histories and Density Matrices}

To understand the application of decoherence at the universal level, and the emergence of branches (branching histories), we can appeal to DH. This is the approach adopted by \citet{saunders1993decoherence, saunders1995time} and \citet{wallace2003everett, Wallace2012}, and has become accepted in much of the contemporary discussion about EQM (\citet{sep-qm-decoherence}, \S3.1; \citet{wilson2020nature}, \S2). We review its essential features and show how they apply equally to WFR and DMR. 

As \citet{GH1994} observed, there are two ways to understand decoherence. The first is the vanishing of the off-diagonal terms in the reduced density matrix of the subsystem, as used in the example above as well as in the derivations of master equations for decoherence (\citet{schlosshauer2007decoherence}). The second, mostly used in approaches such as DH as developed by \citet{gell1990quantum}, quantifies how quasi-classical histories, described by series of projection operators that correspond to branches, emerge from the  universal quantum state and evolve almost independently towards the future, so that probabilities can be consistently assigned to them. The two characterizations are connected, but the second one is especially useful for understanding the emergence of branching histories in EQM.  Even though the division between the system and its environment is readily available in experimental contexts, it is somewhat arbitrary and we should avoid depending on it for justifying the branching structure in EQM. DH is ideal for that purpose, since it applies to prediction in quantum cosmology, the study of the quantum universe as a whole.\footnote{\x{\citet{griffiths1984consistent} and \citet{omnes1988logicala, omnes1988logicalb, omnes1988logicalc, omnes1989logicald} have developed similar approaches. \citet{sep-qm-decoherence} provides a helpful review. For simplicity, we focus on DH as developed by \citet{gell1990quantum}. }}

We review some elements of DH and point out two facts relevant to our project. According to \citet{gell1990quantum}, there are ``three forms of information necessary for prediction in quantum cosmology.'' 
\begin{enumerate}
    \item The quantum state of the universe is given by a density matrix $\rho$. 
    \item Observables (in the Heisenberg picture) are given by projection operators $P(t)$ on Hilbert space.
    \item The projection operators evolve in time by $P(t)= e^{iHt/\hbar}P(0)e^{-iHt/\hbar}$, with $H$ the Hamiltonian. 
\end{enumerate}
In this framework, the universal quantum state is given by a density matrix, of which a pure state is a special case.  As such, it is compatible with DMR\textsubscript{E}. (Indeed, later in their paper \citet{gell1990quantum} consider the special case only to illustrate how Everettian branches emerge in the more familiar setting, an underlying pure state.) 

Notice that the use of a density matrix in the formalism already introduces cross terms in the universal quantum state, even when the state is pure. This can be seen in the simple example of the pure state considered in \S2.1: 
\begin{equation}\label{universal2}
    |S+E\rangle \langle S+E | = c_1^2 |S_1\rangle |E_1\rangle \langle S_1| \langle E_1| + c_2^2 |S_2\rangle |E_2\rangle \langle S_2| \langle E_2| + c.t.
\end{equation}
with $c.t. = c_1 c_2^* |S_1\rangle |E_1\rangle \langle S_2| \langle E_2| + c_2 c_1^* |S_2\rangle |E_2\rangle \langle S_1| \langle E_1|$. People [Removed for review] have worried that any cross term in the universal quantum state corresponds to a branch of inconsistent results (e.g. a device indicating both ``1'' and ``2''), which changes how the Born rule should be understood. That will not be the case, as we explain below.  

The projection operators represent ``yes-no'' observables, such as whether or not the indicator lights up at Dial 6 of some measurement device. At time $t$, \textit{an exhaustive set of mutually exclusive alternatives} can be represented by a set of projection operators $(P_1^k(t), P_2^k(t), ...)$ that sum to  identity and are mutually orthogonal: 
\begin{equation}
    \sum_\alpha P_\alpha^k(t) = \mathbb{I}, 
\end{equation}
\begin{equation}
    P_\alpha^k(t) P_\beta^k(t) = \delta_{\alpha\beta} P_\alpha^k(t).  
\end{equation}
Hence, $\Hilbert_\alpha^k$'s, the subspaces they project onto, provide an orthogonal decomposition of Hilbert space, $\Hilbert=\bigoplus_\alpha \Hilbert_\alpha^k$. The different exhaustive sets of mutually exclusive alternatives can be partially ordered by how coarse-grained they are on Hilbert space. The most coarse-grained one corresponds to the singleton set of the unit operator on Hilbert space, and the most fine-grained ones correspond to the sets of one-dimensional projectors. 

A \textit{history} is a time sequence of alternatives: 
\begin{equation}
    [P_\alpha]=(P_{\alpha_1}^1(t_1), P_{\alpha_2}^2(t_2), ..., P_{\alpha_n}^n(t_n)).
\end{equation}
It corresponds to a time-parametrized path in Hilbert space across various subspaces, starting from $\Hilbert_{\alpha_1}^1$ and ending in $\Hilbert_{\alpha_n}^n$. If the alternative subspaces represent alternative macrostates of distinct pointer readings, it can represent a history where detector 1 indicates ``LEFT'' at $t_1$,  detector 2 indicates ``UP'' at $t_2$, ... and detector n indicates ``YES'' at $t_n$. 

By this construction, since no projection operator corresponds to inconsistent results, there is no single history where at $t_1$ detector 1 both indicates ``LEFT'' and does not indicate ``LEFT.'' Applying this observation to the earlier example,  only the first two terms in (\ref{universal2}) correspond to states of any histories, while the c.t. do not. Hence, c.t. do not correspond to states or measurement outcomes at any time in any history. 

An  important element of DH is the definition of a ``decoherence functional,''  $D([\mathrm{history}'], [\mathrm{history}])$, with respect to the density matrix $\rho$:
\begin{equation}
    D_\rho([P_{\alpha'}], [P_\alpha])= Tr [ P_{\alpha_n'}^n(t_n) ... P_{\alpha_1'}^1(t_1) \rho P_{\alpha_1}^1(t_1) ... P_{\alpha_n}^n(t_n)].
\end{equation}
A set of coarse-grained histories is said to decohere if the cross terms of $D$ (product of two alternative histories acting on the universal quantum state)  are vanishingly small: 
\begin{equation}\label{DHrule}
    D_\rho([P_{\alpha'}], [P_\alpha])\approx 0, \mathrm{for } \, \alpha_k'\neq \alpha_k.
\end{equation}
Such decoherent histories $[P_{\alpha'}]$ and $[P_\alpha]$ arising from $\rho$ will (for a very long time) interfere only negligibly and evolve almost independently, such that the classical probability calculus approximately applies. We may say that branching histories emerge from the universal density matrix. (Typically, the projectors so selected are high-dimensional.)

In DH, the Born rule takes the form that probabilities of sequences of measurement outcomes are the diagonal elements (non-cross terms) of $D$ applied to decoherent histories: 
\begin{equation}\label{DHbornrule}
   Pr([P_\alpha]) = D_\rho([P_{\alpha}], [P_\alpha]).
\end{equation}
This completes our brief summary of DH. 

We emphasize two facts relevant to our project. First,  DH and its explanation of branching histories evidently apply to both WFR and DMR, since it is formulated with a universal density matrix (that may be pure or mixed). Second, in DH, cross terms such as $|S_1\rangle |E_1\rangle \langle S_2| \langle E_2|$ appear in the universal quantum state for both pure states and mixed states, but do not correspond to states $P_{\alpha_k}^k(t_k)$ in any of the decoherent histories $[P_\alpha]$. On DH, possible outcomes of any measurement or possible payoffs of any games correspond to states of some decoherent histories, and different decoherent histories will not recombine (for a very long time). Hence, an Everettian agent who accepts the emergence of branches as decoherent histories will not regard cross terms (in the decoherent bases) as representing possible outcomes of any measurement or possible payoffs of any games. Call this rule \textbf{Ignore Cross Terms}. 

With those understood, we can go back to the more familiar picture of quantum states evolving in time (the Schr\"odinger picture, as used in the discussions below, which is for the purpose at hand equivalent to the Heisenberg picture), with the non-cross terms representing branches at a time, as is standard in typical presentations of measurement situations and the Born rule derivations. DH is in the background, providing a principled reason for why c.t. do not matter for self-location or decision-theoretic considerations: they do not represent states or measurement outcomes in any branches. In the next section, we investigate whether we can extend existing justifications for the Born rule from WFR\textsubscript{E} to DMR\textsubscript{E}. }

\section{The Born Rule}


The motivation for deriving the Born rule is the fact that WFR\textsubscript{E} runs into a \textit{problem of probability}: given that every branch of the wavefunction exists, how can one make sense of quantum-mechanical probabilities? The Born rule tells us that the squared-amplitudes associated with each branch should be interpreted as the probabilities of outcomes on that branch. Intuitively, for an outcome to occur with some probability (that isn't 1) is for it to possibly not occur. In WFR\textsubscript{E}, though, \textit{every} branch -- every possible outcome of a measurement -- always obtains. How, then, can we defend the Born rule and the probabilities it prescribes?  

Consider Alice, an experimenter, performing an $x$-spin measurement on an electron prepared in $\lvert\downarrow_z\rangle$, a state of equal superposition of $x$-spin-up and $x$-spin-down:
\begin{equation}
\lvert\downarrow_z\rangle = \frac{1}{\sqrt{2}}\bigg[\lvert\uparrow_x\rangle - \lvert\downarrow_x\rangle\bigg]
\end{equation}
Then, pre-measurement, the universal wavefunction is in the `ready' state $R$: 
\begin{equation}
    \Psi_R = \lvert\downarrow_z\rangle \lvert R_A \rangle \lvert R_{D_1} \rangle \lvert R_{E} \rangle
\end{equation}
where Alice, $A$, is in the `ready' state $\lvert R_A \rangle$, the measurement device $D_1$ in the state $\lvert R_{D_1} \rangle$ is `ready' to display one of two measurement outcomes $\{\uparrow, \downarrow\}$ , and the environment $E$ (everything else) is `ready' for the measurement by being in the state $\lvert R_{E} \rangle$. After measurement, $\Psi_R$ unitarily evolves into $\Psi_M$:
\begin{equation}
\Psi_M = \frac{1}{\sqrt{2}}\bigg[\lvert\uparrow_x\rangle \lvert {x_\uparrow}_A \rangle \lvert \uparrow_{D_1} \rangle \lvert \uparrow_{E} \rangle - \lvert\downarrow_x\rangle \lvert {x_\downarrow}_A \rangle \lvert \downarrow_{D_1} \rangle \lvert \downarrow_{E} \rangle \bigg]
\end{equation}
with two outcomes with equal amplitudes $\frac{1}{\sqrt{2}}$. The Born rule prescribes that the probability of each outcome occurring is the squared-amplitude (or branch weight) associated with that outcome, i.e. $(\frac{1}{\sqrt 2})^2 = \frac{1}{2}$. 

Much work has been done by defenders of WFR\textsubscript{E} to justify the Born rule. We consider three programs: the Sebens-Carroll program, the McQueen-Vaidman program, and the decision-theoretic program.  We hope to show that \textit{if} we accept any of these programs, as Everettian defenders of Born rule probabilities do, then the justifications for the Born rule can be extended to DMR\textsubscript{E}. 


In what follows, we set aside the well-known problems of circularity surrounding the use of decoherence in the Everettian justification of probability, as we do not offer  new solutions to them. Those problems will affect WFR\textsubscript{E} just as much as DMR\textsubscript{E}. (See \citet{Baker2007} and \citet{kent2010}.) Likewise, we sidestep concerns about the metaphysics of branching:
whether branching occurs `globally' or `locally', which involves the question of whether Alice immediately branches into two copies of themselves after a measurement is made on the electron but before they observe the measurement outcome. While \citet[21]{mcqueenvaidman2018} denies that Alice has branched (hence rejecting `global' branching), Sebens-Carroll's program explicitly depends on `global' branching (\cite[pp. 33--35]{sebenscarroll2016}). (See also \citet{Kent2015} and \citet[\S2.2]{NeyForthcoming-NEYTAF-2}). Since our goal here is \textit{not} to adjudicate between these programs, we won't take a stance. After all, they are both compatible with DH and its justification for \textbf{Ignore Cross Terms}.  We shall show that \textit{both} programs can be readily generalized to DMR\textsubscript{E}.

\subsection{The Sebens-Carroll Program}

\citet{sebenscarroll2016} propose the following strategy: the probabilities ascribed to each branch by the Born rule are to be interpreted as Alice's \textit{self-locating} uncertainty as to which branch they're located in, post-measurement but before they observe the measurement outcome. Their strategy relies on the fact that Alice (i) knows the universal wavefunction, (ii) has undergone branching due to some measurement having been performed,
but (iii) may not be able to discern which branch they're on \textit{prior to observing the measurement outcome} due to each copy of Alice, post-branching, having qualitatively identical internal states as each other. 

In this `post-measurement pre-observation' period, the universal wavefunction is:
\begin{equation}
\Psi_{P} = \frac{1}{\sqrt{2}}\bigg[\lvert\uparrow_x\rangle \lvert R_{A} \rangle \lvert \uparrow_{D_1} \rangle \lvert \uparrow_{E} \rangle + \lvert\downarrow_x\rangle \lvert R_{A} \rangle \lvert \downarrow_{D_1} \rangle \lvert \downarrow_{E} \rangle \bigg]
\end{equation}
That is, while the measurement has been performed and branching has occurred (including for Alice) resulting in two branches associated with $\lvert\uparrow_x\rangle$ and $\lvert\downarrow_x\rangle$ respectively, Alice remains in the `ready' state because they have not observed the measurement outcome yet. They thus have self-locating uncertainty -- subjective credences -- as to which branch of the wavefunction they might be in.

\citet{sebenscarroll2016} propose an intuitive epistemic principle with which they justify Alice's use of the Born rule, where the probabilities are now interpreted in terms of \textit{subjective credences}:
\begin{quote}
    \textbf{Epistemic Separability Principle (ESP):} Suppose that universe $U$ contains within it a set of subsystems $S$ such that every agent in an internally qualitatively identical state to agent $A$ is located in some subsystem which is an element of $S$. The probability that $A$ ought to assign to being located in a particular subsystem $S$ given that they're in $U$ is identical in any possible universe which also contains subsystems $S$ in the same exact states (and does not contain any copies of the agent in an internally qualitatively identical state that are not located in $S$).
    \begin{equation}
    P(X \mid U) = P(X \mid S)
    \end{equation}
\end{quote}
where $P(A\mid B)$ is the conditional probability of $A$ given $B$. An agent, when ascribing credences to each branch, should restrict attention only to those subsystems containing copies of themselves which are internally qualitatively identical. Given ESP, agents ought to ignore everything outside of those subsystems of concern because they are irrelevant to the agent's consideration of credences. In quantum mechanics, the standard way to do that is to construct a density matrix for e.g. $\Psi_P$ and then trace out the irrelevant degrees of freedom, ending up with the relevant reduced density matrix. Supposing that the subsystem which an agent, such as Alice, cares about is the subsystem containing (copies of) Alice and the $n$ measurement outcomes $O_n$ on $D_1$, then the reduced density matrix of interest for Alice is simply $\rho^{AD_1}$ and the credences should be assigned according to:
\begin{equation}
    Pr(O_n \mid \Psi_P) = Pr(O_n \mid \rho^{AD_1})
\end{equation}

\subsubsection{Generalizing the Sebens-Carroll Program to DMR\textsubscript{E}}

We now show how this strategy can be generalized to DMR\textsubscript{E}.



\subsubsection*{Case 1: Pure states}

We start with the simple case of a pure state. Here, the strategy is essentially identical to Seben \& Carroll's for WFR\textsubscript{E} via pure \textit{wavefunctions}, except it's in terms of DMR\textsubscript{E} via pure \textit{density matrices}. 

An observer, Alice, is about to make a $z$-spin measurement of some subsystem, say, an electron prepared in the $x$-spin-down state $\lvert\downarrow_x\rangle$. $D_1$ is in the state $\lvert R_{D_1} \rangle$, ready to show the measurement outcome. Alice is in the state $\lvert R_A \rangle$, ready to observe the measurement outcome. The rest of the universe, i.e. the environment, is also in the `ready' state $\lvert R_{E} \rangle$. 

Given ESP, Alice's self-locating uncertainty should only depend on the subsystems containing Alice and $D_1$: Alice is considering their self-location uncertainty due to branching, occurring as a result of spin measurement, and $D_1$ is the only subsystem showing the outcome of that measurement. Everything else is irrelevant.  

This means that Alice could also consider a second display, $D_2$, likewise in the ready state, represented by $\lvert R_{D_2} \rangle$. The set-up of $D_2$ is irrelevant to Alice's considerations over states of $D_1$, and so they can entertain the possibility of $D_2$ being set up in different configurations without affecting their considerations about self-locating uncertainty concerning measurements on the electron and the outcomes of those measurements displayed on $D_1$.  (Why they would do this will become apparent later.)

Pre-measurement, given the above, the fundamental quantum state of the universe, given by a pure density matrix, can be represented by the `ready' state $\rho_R$: 
\begin{equation}
    \rho_R =    \lvert R_{E} \rangle  \lvert R_{D_2} \rangle \lvert R_{D_1}  \rangle \lvert R_A \rangle \lvert\downarrow_z\rangle \langle\downarrow_z\rvert \langle R_A \rvert \langle R_{D_1} \rvert \langle R_{D_2} \rvert \langle R_{E} \rvert 
\end{equation}
$D_1$ displays the measurement outcome, with two possible outputs $\{\uparrow, \downarrow\}$. That is, it displays $\uparrow$ if the electron was measured to be in the $x$-spin up direction ($\uparrow_x$), and $\downarrow$ if the electron was measured to be in the $x$-spin down direction ($\downarrow_x$). Alice can also suppose that $D_2$, too, has two possible outputs $\{\heartsuit, \diamondsuit\}$ correlated in some way with the outcomes of $D_1$. Again, given ESP, the set-up of $D_2$ per se should be irrelevant to their assignment of self-locating uncertainty given their observation of the measurement outcomes given by $D_1$. Two possible set-ups for $D_2$ can be considered:
\begin{itemize}
    \item \textbf{Set-up $\alpha$}: $D_2$ displays $\heartsuit$ if $D_1$ displays $\uparrow$, and $\diamondsuit$ if the $D_1$ displays $\downarrow$. 
    \item \textbf{Set-up $\beta$}: $D_2$ displays $\diamondsuit$ if $D_1$ displays $\uparrow$, and $\heartsuit$ if the $D_1$ displays $\downarrow$.
\end{itemize}
To set up for our next case, and to make the correlations between the various displays clear for each set-up, we can also write the set-ups as per \textbf{Table 1}.

\begin{table}[h]
\centering
\begin{tabular}{| p{2cm}|p{1cm}|p{1cm}||p{1cm}|p{1cm}|  }
\cline{2-5}
\multicolumn{1}{c|}{} & \multicolumn{4}{|c|}{\textbf{Set-up}} \\
\cline{2-5}
\multicolumn{1}{c|}{} & \multicolumn{2}{|c||}{$\alpha$} & \multicolumn{2}{|c|}{$\beta$} \\
\hline
\hfil Electron & \hfil $\uparrow_x$ & \hfil $\downarrow_x$ & \hfil  $\uparrow_x$  & \hfil $\downarrow_x$ \\
\hline
\hfil $D_1$ & \hfil $\uparrow$ & \hfil $\downarrow$ & \hfil $\uparrow$ & \hfil $\downarrow$ \\
\hline
\hfil $D_2$ & \hfil $\heartsuit$ & \hfil $\diamondsuit$ & \hfil $\diamondsuit$ & \hfil $\heartsuit$ \\
\hline
\end{tabular}
\captionsetup{width=8.2cm}
\caption{Two possible set-ups $\alpha$ and $\beta$, with the only difference (for now) being two possible choices of display output set-ups for $D_2$.} 
\label{tab:fig1}
\end{table}
We suppose that Alice has access to the quantum state, the dynamical laws, and can consider these possible set-ups, but is not immediately aware of, nor affected by, the measurement outcome. At this point, they can consider their self-locating uncertainty. Post-measurement pre-observation of $D_1$, Alice can consider one possibility: $\rho_R$, given set-up $\alpha$, unitarily evolves into
\begin{equation}
\begin{aligned}
    \rho_\alpha = & \frac{1}{2} \bigg[      \lvert \uparrow_{\alpha E} \rangle \lvert \heartsuit_{D_2} \rangle   \lvert \uparrow_{D_1} \rangle  \lvert R_A \rangle \lvert\uparrow_x\rangle \langle\uparrow_x\rvert \langle R_A \rvert \langle \uparrow_{D_1} \rvert \langle \heartsuit_{D_2} \rvert \langle \uparrow_{\alpha E} \rvert \\ 
    & + \lvert \downarrow_{\alpha E} \rangle \lvert \diamondsuit_{D_2} \rangle  \lvert \downarrow_{D_1} \rangle  \lvert R_A \rangle  \lvert \downarrow_x\rangle   \lvert  \langle\downarrow_x\rvert \langle R_A \rvert \langle \downarrow_{D_1} \rvert \langle \diamondsuit_{D_2} \rvert \langle \downarrow_{\alpha E} \rvert \bigg]  \y{+ c.t.}
\end{aligned}
\end{equation}
\noindent \x{This universal state contains two branches, as the c.t. do not correspond to any branch (by \textbf{Ignore Cross Terms}).} For notational convenience, we rewrite $\rho_\alpha$ as: 
\begin{equation}\label{rhoalpha}
\begin{aligned}
    \rho_\alpha = & \frac{1}{2} \bigg[ \rho_{\uparrow_x} \rho_{R_A} \rho_{\uparrow_{D_1}} \rho_{\heartsuit_{D_2}} \rho_{\uparrow_{\alpha E}} \\ 
    & + \rho_{\downarrow_x} \rho_{R_A} \rho_{\downarrow_{D_1}} \rho_{\diamondsuit_{D_2}} \rho_{\downarrow_{\alpha E}}  \bigg]  \y{+ c.t.}
\end{aligned}
\end{equation}
\noindent But they could have considered set-up $\beta$ instead. If that were the case, $\rho_R$ would have instead unitarily evolved into 
\begin{equation}\label{rhobeta}
\begin{aligned}
    \rho_\beta = & \frac{1}{2} \bigg[ \rho_{\uparrow_x} \rho_{R_A} \rho_{\uparrow_{D_1}} \rho_{\diamondsuit_{D_2}} \rho_{\uparrow_{\beta E}} \\ 
    & + \rho_{\downarrow_x} \rho_{R_A} \rho_{\downarrow_{D_1}} \rho_{\heartsuit_{D_2}} \rho_{\downarrow_{\beta E}}  \bigg]
     \y{+ c.t.}
\end{aligned}
\end{equation}
To emphasize, these are \textit{not} the only configurations $D_2$ can have, but rather two \textit{possible} set-ups that are available to use for our derivation. We're allowed to consider these configurations since ESP asks Alice to restrict their attention only to the subsystems containing $A$ and $D_1$; possible changes in everything else which do not affect $A$ and $D_1$ per se can be entertained without affecting Alice's considerations about their self-locating uncertainty. 

In particular, note that when we trace out these irrelevant degrees of freedom (by the lights of ESP) from $\rho_\alpha$ and $\rho_\beta$, the resultant reduced density matrices are equivalent:
\begin{equation}\label{ESPrhoAD1}
\begin{aligned}
    \rho^{AD_1} = \rho_\alpha^{AD_1} & = \rho_\beta^{AD_1} \\
    & \ec{\approx} \frac{1}{2}\bigg[\rho_{R_A} \rho_{\uparrow_{D_1}} + \rho_{R_A} \rho_{\downarrow_{D_1}}\bigg]
\end{aligned} 
\end{equation}
Two branches emerge as a result of decoherence -- associated with definite measurement outcomes for states $\uparrow_x$ and $\downarrow_x$. We can associate each branch of $\rho_{\alpha}$ and $\rho_{\beta}$ with each column of \textbf{Table 1}.
Now, the goal is to show, given ESP, that Alice ought to take each branch of $\rho^{AD_1}$ to be equiprobable: Alice ought to assign credences $Pr(\uparrow \;\mid \rho_\alpha) = Pr(\downarrow \;\mid \rho_\alpha) = 1/2$.

Since $\rho_\alpha^{AD_1} = \rho_\beta^{AD_1}$ as per (\ref{ESPrhoAD1}):
\begin{equation}\label{purestateproof1}
    Pr(\uparrow \; \mid \rho_\alpha) = Pr(\uparrow \; \mid \rho_\beta)
\end{equation}
We can also use ESP to restrict attention to subsystems containing (different possible) $D_2$ and Alice, if we wanted to consider Alice's self-locating uncertainty over being in a branch correlated with an outcome of $D_2$. By tracing out $D_1$, the electron state, and the rest of the environment, we see that
\begin{equation}\label{ESPrhoAD2}
\begin{aligned}
    \rho^{AD_2} = \rho_\alpha^{AD_2} & = \rho_\beta^{AD_2} \\
    & \ec{\approx} \frac{1}{2}\bigg[\rho_{R_A} \rho_{\heartsuit_{D_2}} + \rho_{R_A} \rho_{\diamondsuit_{D_2}}\bigg]
\end{aligned}
\end{equation}
Hence, given ESP, the credences that Alice ascribes to outcomes of $D_2$, e.g. $\diamondsuit$, in both set-ups $\alpha$ and $\beta$ should also be identical:
\begin{equation}\label{purestateproof2}
    Pr(\diamondsuit \mid \alpha) = Pr(\diamondsuit \mid \beta)
\end{equation}
Note from \textbf{Table \ref{tab:fig1}}, or from (\ref{rhoalpha}), that the $\downarrow$-branch \textit{just is} the $\diamondsuit$-branch in a universe with the quantum state $\rho_\alpha$, that is, a universe where set-up $\alpha$ was implemented. So Alice's self-locating uncertainty about being in the $\downarrow$-branch must be the same as that for being in the $\diamondsuit$-branch. Alice knows this \textit{same-branch relationship} since, \textit{ex hypothesi}, they have access to the quantum state. Hence, they can use this to conclude that:
\begin{equation}\label{purestateproof3}
    Pr(\downarrow \; \mid \alpha) = Pr(\diamondsuit \mid \alpha)
\end{equation}
\noindent Likewise, for a universe with the quantum state $\rho_\beta$, Alice can observe that the $\uparrow$-branch \textit{just is} the $\diamondsuit$-branch. Hence:
\begin{equation}\label{purestateproof4}
    Pr(\uparrow \; \mid \beta) = Pr(\diamondsuit \mid \beta)
\end{equation}
\noindent Therefore, putting (\ref{purestateproof1}), (\ref{purestateproof2}), (\ref{purestateproof3}), and (\ref{purestateproof4}) together, we see that:
\begin{equation}
\begin{aligned}\label{purestateproofresult}
    & 1. \; Pr(\uparrow \; \mid \alpha) = Pr(\uparrow \; \mid \beta) \;\;\;\;\;\; \text{from } (\ref{purestateproof1})\\
    & 2. \;  Pr(\uparrow \; \mid \beta) = Pr(\diamondsuit \mid \beta) \;\;\;\;\; \text{from } (\ref{purestateproof4})\\
    & 3. \;  Pr(\diamondsuit \mid \beta) = Pr(\diamondsuit \mid \alpha) \;\;\;\; \text{from } (\ref{purestateproof2}) \\
    & 4. \;  Pr(\diamondsuit \mid \alpha) = Pr(\downarrow \; \mid \alpha) \;\;\;\;\; \text{from } (\ref{purestateproof3}) \\
    & \therefore Pr(\uparrow \; \mid \alpha) = Pr(\downarrow \; \mid \alpha)
\end{aligned}
\end{equation}
From (\ref{purestateproofresult}), we see that considerations of the probabilities prescribed by ESP require Alice to assign equal credences, when considering self-locating uncertainty, to both the $\uparrow$-branch and the $\downarrow$-branch. This uniquely determines their credences for being in either branch to be equal to that branch's weight, $1/2$. 

This vindicates the Born rule for DMR\textsubscript{E} for the simple case of \textit{equal-weight} superpositions represented by pure-state density matrices.  


\subsubsection*{Case 2: Mixed states}

Suppose the universal state is a mixture of two pure density matrices. Would Sebens \& Carroll's proof work then? We show that it can be done, by working through an explicit case and  providing an algorithm for generalizing this to arbitrary density matrices. 

Let $\rho_{\downarrow_z}$ be the pure density matrix representing an electron in the $\downarrow_z$ state, and let $\rho_{\downarrow_x}$ be the pure density matrix representing an electron in the $\downarrow_x$ state. Then, suppose Alice is in a universe in the mixed state: 
\begin{equation}
    \rho_{R^\prime} = \frac{1}{2}\bigg(\rho_{\downarrow_z} \rho_{R_A} \rho_{R_{D_1}} \rho_{R_{E}} + \rho_{\downarrow_x} \rho_{R_A} \rho_{R_{D_1}} \rho_{R_{E}}\bigg)
\end{equation}
Depending on the environment, especially the measurement device being used, $\rho_{R^\prime}$ will evolve differently given DMR\textsubscript{E}, just as with WFR\textsubscript{E}. 

Suppose we made a measurement for $x$-spin.
Then $\rho_{R^\prime}$ will unitarily evolve into: \begin{equation}
    \rho_{M^\prime} = \frac{1}{4}\rho_{\uparrow_x} \rho_{R_A} \rho_{\uparrow_{D_1}} \rho_{\uparrow_{E}} +  \frac{3}{4}\rho_{\downarrow_x} \rho_{R_A} \rho_{\downarrow_{D_1}} \rho_{\downarrow_{E}} \ec{+ c.t.}
\end{equation}
\x{Once again, this universal state contains two branches, as the c.t. do not correspond to any branch (by \textbf{Ignore Cross Terms}).} Now, to determine Alice's self-locating uncertainty over the possible branches of $\rho_{M^\prime}$, we consider two possible scenarios, $\mu$ and $\nu$, in which additional displays in the environment, $D_2$ with associated outputs $\{\heartsuit, \diamondsuit\}$, $D_3$ with $\{\clubsuit, \spadesuit\}$, and $D_4$ with $\{\cross, \star\}$, may display results (see Table 2). 
\begin{table}[h]
\centering
\begin{tabular}{|p{2cm}|p{0.5cm}|p{0.5cm}|p{0.5cm}|p{0.5cm}||p{0.5cm}|p{0.5cm}|p{0.5cm}|p{0.5cm}|  }
\cline{2-9}
\multicolumn{1}{c|}{} & \multicolumn{8}{|c|}{\textbf{Set-up}} \\
\cline{2-9}
\multicolumn{1}{c|}{} & \multicolumn{4}{|c||}{$\mu$} & \multicolumn{4}{|c|}{$\nu$} \\
\hline
\hfil Electron & \hfil $\uparrow_x$ & \hfil $\downarrow_x$ & \hfil  $\downarrow_x$  & \hfil $\downarrow_x$ & \hfil $\uparrow_x$ & \hfil $\downarrow_x$ & \hfil  $\downarrow_x$  & \hfil $\downarrow_x$ \\
\hline
\hfil $D_1$ & \hfil $\uparrow$ & \hfil $\downarrow$ & \hfil $\downarrow$ & \hfil $\downarrow$ & \hfil $\uparrow$ & \hfil $\downarrow$ & \hfil $\downarrow$ & \hfil $\downarrow$ \\
\hline
\hfil $D_2$ & \hfil $\diamondsuit$ & \hfil $\heartsuit$ & \hfil $\diamondsuit$ & \hfil $\diamondsuit$ & \hfil $\heartsuit$ & \hfil $\diamondsuit$ & \hfil $\diamondsuit$ & \hfil $\diamondsuit$ \\
\hline
\hfil $D_3$ & \hfil $\clubsuit$ & \hfil $\clubsuit$ & \hfil $\spadesuit$ & \hfil $\clubsuit$ & \hfil $\spadesuit$ & \hfil $\clubsuit$ & \hfil $\clubsuit$ & \hfil $\clubsuit$ \\
\hline
\hfil $D_4$ & \hfil $\star$ & \hfil $\star$ & \hfil $\star$ & \hfil $\cross$ & \hfil $\cross$ & \hfil $\star$ & \hfil $\star$ & \hfil $\star$ \\
\hline
\end{tabular}
\captionsetup{width=9.8cm}
\caption{Two possible set-ups $\mu$ and $\nu$, corresponding to two possible choices of display output set-ups for $D_2$, $D_3$, and $D_4$.} 
\label{tab:fig2}
\end{table}

 There are many physically possible ways to achieve the above correlations by performing transformations on the environment (\cite[p.46]{sebenscarroll2016}). For example, someone could conditionally measure a second particle upon observing $D_1$'s display, and $D_2$ could conditionally display the outcome of that measurement instead. For instance, in set-up $\mu$, $D_2$/$D_3$/$D_4$ could display $\diamondsuit$/$\clubsuit$/$\star$ if $D_1$ displays $\uparrow$. If $D_1$ displays $\downarrow$, then the other displays will display the result of some measurement on a second particle which yields three distinct outcomes, only showing $\spadesuit$/$\cross$/$\heartsuit$ on one of the outcomes. On set-up $\nu$, $D_2$, $D_3$ and $D_4$ might just output $\heartsuit$/$\spadesuit$/$\cross$ if $D_1$ displays $\uparrow$, and $\diamondsuit$/$\clubsuit$/$\star$ if $D_1$ displays $\downarrow$. 

As with \textbf{Case 1}, we can associate each column of a set-up in \textbf{Table 2} with a decohered branch. Note that $\uparrow$, $\heartsuit$, $\spadesuit$, and $\cross$ \textit{each uniquely picks out a branch} in set-up $\mu$, and that these four symbols also \textit{all pick out the same branch} in $\nu$. As with \textbf{Case 1}, we'll use these facts to derive the Born rule probabilities from the universal density matrix.

Corresponding to each possible set-up, $\rho_{M^\prime}$ could have unitarily evolved into two possible states: 
\begin{equation}
\begin{aligned}
    \rho_\mu &  =  \frac{1}{4}  \rho_{\uparrow_x} \rho_{R_A} \rho_{\uparrow_{D_1}} \rho_{\diamondsuit_{D_2}} \rho_{\clubsuit_{D_3}} \rho_{\star_{D_4}} \rho_{\mu1_E} \\
    & + \frac{1}{4} \rho_{\downarrow_x} \rho_{R_A} \rho_{\downarrow_{D_1}} \rho_{\heartsuit_{D_2}} \rho_{\clubsuit_{D_3}} \rho_{\star_{D_4}} \rho_{\mu2_E} \\
    & + \frac{1}{4} \rho_{\downarrow_x} \rho_{R_A} \rho_{\downarrow_{D_1}} \rho_{\diamondsuit_{D_2}} \rho_{\spadesuit_{D_3}} \rho_{\star_{D_4}} \rho_{\mu3_E} \\
    & + \frac{1}{4} \rho_{\downarrow_x} \rho_{R_A} \rho_{\downarrow_{D_1}} \rho_{\diamondsuit_{D_2}} \rho_{\clubsuit_{D_3}} \rho_{\cross_{D_4}} \rho_{\mu4_E} \\
    & \ec{+ c.t.}
\end{aligned}
\end{equation}
Or:
\begin{equation}
\begin{aligned}
    \rho_\nu & =  \frac{1}{4} \rho_{\uparrow_x} \rho_{R_A} \rho_{\uparrow_{D_1}} \rho_{\heartsuit_{D_2}} \rho_{\spadesuit_{D_3}} \rho_{\cross_{D_4}} \rho_{\mu1_E} \\
    & + \frac{1}{4} \rho_{\downarrow_x} \rho_{R_A} \rho_{\downarrow_{D_1}} \rho_{\diamondsuit_{D_2}} \rho_{\clubsuit_{D_3}} \rho_{\star_{D_4}} \rho_{\mu2_E} \\
    & + \frac{1}{4} \rho_{\downarrow_x} \rho_{R_A} \rho_{\downarrow_{D_1}} \rho_{\diamondsuit_{D_2}} \rho_{\clubsuit_{D_3}} \rho_{\star_{D_4}} \rho_{\mu3_E} \\
    & + \frac{1}{4} \rho_{\downarrow_x} \rho_{R_A} \rho_{\downarrow_{D_1}} \rho_{\diamondsuit_{D_2}} \rho_{\clubsuit_{D_3}} \rho_{\star_{D_4}} \rho_{\mu4_E} \\
    & \ec{+ c.t.}
\end{aligned}
\end{equation}
We're now able to derive our main result. Using ESP again, we see that:
\begin{equation}\label{eq:47}
    Pr(\uparrow \; \mid \mu) =  Pr(\uparrow \; \mid \nu)
\end{equation}
\begin{equation}\label{eq:48}
    Pr(\heartsuit\mid \mu) =  Pr(\heartsuit \mid \nu)
\end{equation}
Furthermore, by scrutinizing $\rho_\mu$ and $\rho_\nu$, or by consulting \textbf{Table 2}, we see that the $\uparrow$-branch just is the $\heartsuit$-branch in $\rho_\nu$ and hence:
\begin{equation}\label{eq:49}
    Pr(\heartsuit \mid \nu) =  Pr(\uparrow \; \mid \nu)
\end{equation}
We've established the equivalence of probabilities for the $\uparrow$-branch and $\heartsuit$-branch in $\rho_\mu$. We do the same for the two remaining branches using the same strategy of using ESP and consulting \textbf{Table 2} to observe same-branch relationships between the symbols. We get: \begin{equation}\label{eq:50}
   Pr(\spadesuit \mid \mu) =  Pr(\spadesuit \mid \nu) = Pr(\uparrow \; \mid \nu)
\end{equation}
\begin{equation}\label{eq:51}
    Pr(\cross \mid \mu) =  Pr(\cross \mid \nu) = Pr(\uparrow \; \mid \nu)
\end{equation}
Hence, from \eqref{eq:47} to \eqref{eq:51}: 
\begin{equation}\label{eq:52}
    Pr(\uparrow \; \mid \mu) = Pr(\heartsuit \mid \mu) = Pr(\spadesuit \mid \mu) = Pr(\cross \mid \mu)
\end{equation}
Since the $\uparrow$-branch, $\heartsuit$-branch, $\spadesuit$-branch, and $\cross$-branch exhaust the branches of $\rho_\mu$ and are mutually exclusive after decoherence, and since they're equiprobable from \eqref{eq:52}, Alice should assign equal credences to being in any of the branches. That is: 
\begin{equation}
    Pr(\uparrow \; \mid \mu) = Pr(\heartsuit \mid \mu) = Pr(\spadesuit \mid \mu) = Pr(\cross \mid \mu) = \frac{1}{4}
\end{equation}
Consulting \textbf{Table 2} again reveals that the $\heartsuit$-branch, $\spadesuit$-branch, and $\cross$-branch are all $\downarrow$-branches. Since they're approximately mutually exclusive (per decoherence) and exhaust all possible branches in which $\downarrow$ shows up in $\rho_\mu$, we have: 
\begin{equation}
    Pr(\downarrow \; \mid \mu) = (Pr(\heartsuit \mid \mu) + Pr(\spadesuit \mid \mu) + Pr(\cross \mid \mu)
\end{equation}
Hence, if Alice ought to assign equal credences of $1/4$ each to $Pr(\heartsuit \mid \mu)$, $Pr(\spadesuit \mid \mu)$, and $Pr(\cross \mid \mu)$, then:
\begin{equation}
    Pr(\downarrow \; \mid \mu) = \frac{3}{4}
\end{equation}
But since Alice rationally ought to assign 1/4 to $Pr(\uparrow \; \mid \mu)$ and 3/4 to $Pr(\downarrow \; \mid \mu)$, Alice rationally ought to follow the Born rule. $\square$

\subsubsection{General Strategy}

Sebens and Carroll's strategy exploits the fact that there are many physically possible and convenient set-ups, i.e. possible environments, such that one can write down the quantum state as a sum of equal-amplitude, and hence equally weighted, branches. They discuss this general proof obliquely in their appendix, but we think their strategy can be explicated much more clearly, especially with the schematic tables we've used. 

For a quantum state that one wishes to split into $N$ equal-amplitude branches, one considers, beyond $D_0$ which displays the original measurement outcome, $N$ further displays in the environment, each of which displays two outputs $\{\star_N, \star_N^\prime\}$. The agent then considers cases in which the $\star_N^\prime$ symbols only show up \textit{once} for the $N^{th}$ display. Per ESP, one may unitarily transform these displays (and systems whose measurement outcomes they represent) in many physically possible ways without affecting one's self-location uncertainty regarding the subsystem containing the agent and $D_0$. Then, the Sebens-Carroll strategy for setting up the two set-ups can be made more explicit in terms of two simple strategies:
\begin{itemize}
    \item \textbf{Diagonalization}: the first set-up diagonalizes, for the $N^{th}$ display, the $\star_N^\prime$ symbol (as seen in \textbf{Table 3} as set-up $\alpha$).
    \item \textbf{Same-Branch}: the second set-up considers a possibility in which the $\star_N^\prime$ symbols all show up in the same branch (seen in \textbf{Table 3} as set-up $\beta$). 
\end{itemize}
Suppose Alice is considering a measurement on some system with $k$ unequally weighted outcomes, $O_k$, such that $D_0$ displays $\{``1", ``2", ... ``k"$\}. Given this set-up, one can consider the schematic \textbf{Table 3}.
\begin{table}[h]
\centering
\begin{tabular}{|p{1.4cm}|p{0.6cm}|p{0.6cm}|p{0.6cm}|p{0.6cm}|p{0.6cm}||p{0.6cm}|p{0.6cm}|p{0.6cm}|p{0.6cm}|p{0.6cm}|}
\cline{2-11}
\multicolumn{1}{c|}{} & \multicolumn{10}{|c|}{\textbf{Set-up}} \\
\cline{2-11}
\multicolumn{1}{c|}{} & \multicolumn{5}{|c||}{$\alpha$} & \multicolumn{5}{|c|}{$\beta$} \\
\hline
\hfil System & \hfil $O_1$ & \hfil $O_1$ & \hfil $O_2$  & \hfil \ldots & \hfil $O_k$ & \hfil $O_1$ & \hfil $O_1$ & \hfil $O_2$  & \hfil \ldots & \hfil $O_k$ \\
\hline
\hfil $D_0$ & \hfil $``1"$ & \hfil $``1"$ & \hfil $``2"$  & \hfil \ldots & \hfil $``k"$ & \hfil $``1"$ & \hfil $``1"$ & \hfil $``2"$  & \hfil \ldots & \hfil $``k"$ \\
\hline
\hfil $D_1$ & \hfil \cellcolor{gray!10}$\star_1^\prime$ & \hfil $\star_1$ & \hfil $\star_1$  & \hfil $\star_1$ & \hfil $\star_1$ & \hfil \cellcolor{gray!10}$\star_1^\prime$ & \hfil $\star_1$ & \hfil $\star_1$ & \hfil $\star_1$ & \hfil $\star_1$\\
\hline
\hfil \ldots & \hfil $\star_2$ & \hfil \cellcolor{gray!10}$\star_2^\prime$  & \hfil $\star_2$  & \hfil $\star_2$ & \hfil $\star_2$ & \cellcolor{gray!10}\hfil $\star_2^\prime$ & \hfil $\star_2$ & \hfil$\star_2$ & \hfil$\star_2$ & \hfil $\star_2$\\
\hline
\hfil \ldots & \hfil \ldots & \hfil \ldots & \hfil \cellcolor{gray!10}\ldots & \hfil \ldots & \hfil\ldots & \hfil \cellcolor{gray!10}\ldots & \hfil \ldots & \hfil \ldots & \hfil \ldots & \hfil \ldots\\
\hline 
\hfil \ldots & \hfil  \ldots  & \hfil \ldots  & \hfil \ldots  & \hfil \cellcolor{gray!10}\ldots  & \hfil \ldots  & \hfil \cellcolor{gray!10}\ldots  & \hfil \ldots & \hfil \ldots  & \hfil \ldots  & \hfil \ldots \\
\hline
\hfil $D_{N}$ & \hfil $\star_{N}$ & \hfil $\star_{N}$ & \hfil $\star_{N}$ & \hfil $\star_{N}$ & \hfil \cellcolor{gray!10} $\star_N^\prime$ & \hfil \cellcolor{gray!10} $\star_N^\prime$ & \hfil $\star_{N}$ & \hfil $\star_{N}$ & \hfil $\star_{N}$ & \hfil $\star_{N}$ \\
\hline
\end{tabular}
\captionsetup{width=11cm}
\caption{Two possible set-ups $\alpha$ and $\beta$, with two possible choices of display output set-ups for each $D_1, D_2, ...,D_N$, with two possible outputs $\star_N$ and $\star^\prime_N$ each. Generally there can be many equal-amplitude branches with the same outcomes (e.g. $O_1$). } 
\label{tab:fig5}
\end{table}

Again, one can generically treat each column of each set-up as an equal-amplitude weighted branch of some possible universal quantum state with set-up $\alpha$ or $\beta$ respectively. Then, one simply uses ESP to judge that, for each $N$:
\begin{equation}
\begin{aligned}
    Pr(\star_N^\prime \mid \alpha) = Pr(\star_N^\prime \mid \beta) 
\end{aligned}
\end{equation}
and consider the same-branch relationships between the symbols in $\rho_\beta$:
\begin{equation}
    Pr(\star_1^\prime \mid \beta) = Pr(\star_2^\prime \mid \beta) = ... = Pr(\star_N^\prime \mid \beta)
\end{equation}
This straightforwardly entails 
\begin{equation}
    Pr(\star_1^\prime \mid \alpha) = Pr(\star_2^\prime \mid \alpha) = ... = Pr(\star_N^\prime \mid \alpha)
\end{equation}
which entails that an agent ought to assign equal credences that they might be located on each of the $N^{th}$ equally weighted branches. But that is the Born rule. 

This is the strategy provided by \citet{sebenscarroll2016}. However, we've clarified the reasoning behind the set-ups by explicitly highlighting \textbf{Diagonalization} and \textbf{Same-Branch} as principles for choosing the appropriate set-ups $\alpha$ and $\beta$. The Sebens-Carroll program readily generalizes to DMR\textsubscript{E}. 

\subsection{The McQueen-Vaidman Program}

Similar to the Sebens-Carroll program, \citet{mcqueenvaidman2018} propose an interpretation of the Born rule in terms of self-locating uncertainty, following earlier attempts by e.g. \citet{Vaidman1998} and \citet{Tappenden2011}. McQueen and Vaidman's setup depends on the fiction of a sleeping pill, which induces the same post-measurement pre-observation uncertainty as the Sebens-Carroll program. Instead of relying on ESP, they use three physical principles:
\begin{itemize}
    \item \textbf{Symmetry:} Symmetric situations should be assigned equal probabilities.
    \item \textbf{No-FTL:} Faster-than-light signaling is impossible; the probability of finding a particle in some location with some state cannot be influenced by actions occurring remotely. 
    \item \textbf{Locality:} The probability of finding a particle somewhere in some state depends only on that particle's quantum state. 
\end{itemize}
Consider a particle described by a subsystem wavefunction with an equal superposition of $N$ very well-localized and remote wave-packets $\lvert L_N \rangle$ each corresponding to the particle being found at the $N$\textsuperscript{th} location, a case of perfect symmetry. $N$ identical measurement apparatus are set up at each of $N$ identically built space-stations, each containing an agent $A_N$ which, for all practical purposes, are identical to one another. Each space-station is located on the circumference of a perfect circle such that the particle has $N$-fold spherical symmetry. For instance, if $N = 3$, then the particle's subsystem wavefunction is described by:
\begin{equation}\label{symmetric-vaidman}
    \Psi_S = \frac{1}{\sqrt{3}}\bigg[\lvert L_1 \rangle + \lvert L_2 \rangle + \lvert L_3 \rangle \bigg]
\end{equation}
Given this situation, each agent -- well aware of the symmetry of the situation -- is put in a sleeping pill situation: they're put to sleep before measurement in a `ready' room, and then moved to a `found' room -- stipulated to be internally identical as the `ready' room -- if the particle is found by the measurement apparatus in their space-station. The measurement then takes place.

Now, given WFR\textsubscript{E}, the agent knows that they will branch into two copies upon looking at the measurement outcome, one remaining in the `ready' room, and one moving to the `found' room. What is the self-locating uncertainty they should ascribe to being in the `found' room? Agent $A_1$ knows that if they're in the `found' room, then the other agents ($A_2$, $A_3$...) are in the `ready' room. But they also know this is true for each other agent: if $A_2$ is in the `found' room, then the others are in the `ready' room, and likewise for $A_3$, $A_4$... $A_N$. This exhausts all the possibilities given the form of the particle's wavefunction. Since each outcome is symmetric given the set-up, under \textbf{Symmetry}, each agent should rationally assign each outcome the same credence. The unique way to assign each outcome a probability is to assign each outcome $\frac{1}{N}$. This corresponds to the squared-amplitude weights of each outcome, and so vindicates the Born rule for this specific symmetric case.  

For asymmetric cases, we keep the same symmetric set-up as before. However,  \textbf{No-FTL} and \textbf{Locality} ensures that changes to the wave-packets at $L_2$, $L_3$, ... of a wavefunction with $N$ remote wave-packets do not influence the wave-packet at $L_1$ in terms of the credences $A_1$ ought to assign to their local measurement outcomes. Then, even if the wavefunction in question evolves from e.g. (\ref{symmetric-vaidman}) to: 
\begin{equation}
    \frac{1}{\sqrt{3}}\lvert L_1 \rangle + \frac{2}{\sqrt{3}}\lvert ? \rangle
\end{equation}
by e.g. transforming the $ \lvert L_2 \rangle$ and $\lvert L_3 \rangle $ wave packets into some arbitrary state $\lvert ? \rangle$, the agent $A_1$ at $L_1$ should assign probabilities \textit{as though} they were in the symmetric case since they have no access to the information that the transformation took place. That is, they should assign 1/3 to the outcome that they find the particle in $L_1$ (and hence end up in the `found' room after awakening from the sleeping pill). 


\subsubsection{Generalizing the McQueen-Vaidman Program to DMR\textsubscript{E}}

\ec{We now generalize the McQueen-Vaidman program to DMR\textsubscript{E}.} 

To begin, the principles of \textbf{Symmetry}, \textbf{No-FTL}, and \textbf{Locality} do not turn on the quantum state's purity. Furthermore, it seems to us that an agent living in a mixed-state multiverse will equally be able to entertain questions about self-locating uncertainty using McQueen \& Vaidman's strategy.

Consider a base case of perfect symmetry. We work out the $N = 3$ case as McQueen and Vaidman (2018) does. Consider, again, $3$ identical measurement apparatus, $M_1$, $M_2$, $M_3$, set up at each of $3$ identically built space-stations, each containing an agent $A_1$, $A_2$, $A_3$ respectively, which, for all practical purposes, are identical to one another. Each space-station is located at locations $1$, $2$, or $3$, on the circumference of a perfect circle such that the particle has $3$-fold spherical symmetry. The agents, measurement devices, and environment $E$ are all in the ready state $R$. There is a particle described in terms of the equal sum of $3$ mixed-state density matrices $\rho_{L_1}$, $\rho_{L_2}$, and $\rho_{L_3}$, corresponding to the particle being well-localized at locations 1, 2, and 3 respectively. So, the universal density matrix, in the ready state, $\rho_R$ is described by:
\begin{equation}
\begin{aligned}
    \rho_R = & \frac{1}{3}\bigg[\rho_{L_1} + \rho_{L_2} + \rho_{L_3} \bigg] \rho_{R_{A_1}} \rho_{R_{A_2}} \rho_{R_{A_3}} \rho_{R_{M_1}} \rho_{R_{M_2}} \rho_{R_{M_3}} \rho_{{E}} 
\end{aligned}
\end{equation}
Each agent -- well aware of the symmetry of the situation -- is put in a sleeping pill situation: they're put to sleep before measurement in a `ready' room, and then moved to a `found' room -- stipulated to be internally identical as the `ready' room -- if the particle is found by the measurement apparatus in their space-station. 

Given DMR\textsubscript{E}, decoherence occurs just like in WFR\textsubscript{E}. The post-measurement universal density matrix evolves into\footnote{\x{This universal state can contain cross terms if expressed in other bases, but such terms do not represent branches by \textbf{Ignore Cross Terms}.}}:
\begin{equation}
\begin{aligned}
    \rho_P = & \frac{1}{3}\bigg[\rho_{L_1} \rho_{R_{A_1}} \rho_{R_{A_2}} \rho_{R_{A_3}} \rho_{\checkmark_{M_1}} \rho_{R_{M_2}} \rho_{R_{M_3}} \rho_{{E_1}} \\
     & + \rho_{L_2} \rho_{R_{A_1}} \rho_{R_{A_2}} \rho_{R_{A_3}} \rho_{R_{M_1}} \rho_{\checkmark_{M_2}} \rho_{R_{M_3}} \rho_{{E_2}} \\
     & + \rho_{L_3} \rho_{R_{A_1}} \rho_{R_{A_2}} \rho_{R_{A_3}} \rho_{R_{M_1}} \rho_{R_{M_2}} \rho_{R_{\checkmark_3}} \rho_{{E_3}} \bigg] \\
\end{aligned}
\end{equation}
\ec{and the reduced density matrix corresponding to the subsystems of interest $S$ is}:
\begin{equation}
\begin{aligned}
    \rho_S \approx & \frac{1}{3}\bigg[\rho_{L_1} \rho_{R_{A_1}} \rho_{R_{A_2}} \rho_{R_{A_3}} \rho_{\checkmark_{M_1}} \rho_{R_{M_2}} \rho_{R_{M_3}} \\
     & + \rho_{L_2} \rho_{R_{A_1}} \rho_{R_{A_2}} \rho_{R_{A_3}} \rho_{R_{M_1}} \rho_{\checkmark_{M_2}} \rho_{R_{M_3}} \\
     & + \rho_{L_3} \rho_{R_{A_1}} \rho_{R_{A_2}} \rho_{R_{A_3}} \rho_{R_{M_1}} \rho_{R_{M_2}} \rho_{R_{\checkmark_3}} \bigg]
\end{aligned}
\end{equation}
Post-measurement pre-observation, each agent, remaining in the $R$ state because they have yet to find out whether they're in the `found' or `ready' room, may ask: What is the self-locating uncertainty they should ascribe to being in the `found' room? Agent $A_1$ knows that if they're in the `found' room, then the other agents ($A_2$, $A_3$...) are in the `ready' room. But they also know this is true for each other agent: if $A_2$ is in the `found' room, then the others are in the `ready' room, and likewise for $A_3$, $A_4$... $A_N$. This exhausts all the possibilities given the form of the particle's wavefunction. Since each outcome is symmetric given the set-up, under \textbf{Symmetry}, each agent should rationally assign each outcome the same credence. The unique way to assign each outcome a probability is to assign each outcome $1/3$. 

Furthermore, each agent may, even in DMR\textsubscript{E}, employ \textbf{No-FTL} and \textbf{Locality}, to entertain the possibility that any asymmetric case can be transformed into the symmetric case without their knowledge -- the same strategy for WFR\textsubscript{E} generalizes to DMR\textsubscript{E}. In summary, the McQueen-Vaidman program does not depend essentially on WFR\textsubscript{E}, and can be readily generalized to DMR\textsubscript{E}. 

\subsection{The Decision-Theoretic Program}

Finally, we turn to the decision-theoretic program. 
The idea is that a rational agent betting on outcomes of measurements for some wavefunction ought to bet in such a way that their credences over these outcomes are governed by the squared-amplitudes of the wavefunction. \citet{deutsch1999} provided the earliest and a particularly simple proof for this result. Interestingly, Deutsch remarked (p. 3136) that ``further generalisation [of his proof] to exotic situations in which the universe as a whole may be in a mixed state is left as an exercise for the reader.''  In what follows, we carry out the exercise and generalize his proof to DMR\textsubscript{E}.\footnote{\x{Two remarks: (1) We are aware of the various issues faced by proposed solutions to the problem of probability. We don't intend to defend Deutsch's proof -- we simply show how the proof can be generalized to DMR. We leave it open that Deutsch's proof -- indeed, any program mentioned here -- require further defenses. (2) The proof in \citet{Wallace2012}'s monograph is another sophisticated (book-length!) rendition of the general decision-theoretic strategy. However, because of space constraints, we leave its generalization to future work.}}

We assume, with Deutsch, some standard non-probabilistic features of quantum mechanics. First, we posit the existence of non-degenerate observables $\hat{X}$, with eigenstates of eigenvalues $x_a$ and associated projection operators $\hat{P}_{x_a}$, and quantum states given by density density matrices $\rho(t)$. Second, we assume the meaningfulness of the notion of outcomes that ``will happen" (rather than occurring with some non-trivial probability, hence avoiding the need for the Born rule which we seek to prove): if system is in eigenstate $|x\rangle\langle x|$ of $\hat{X}$ and $\hat{X}$ is measured, the outcome \textit{will} correspond to eigenvalue $x$. Third, if $\hat{X}$ is measured when system is in arbitrary (pure for Deutsch, mixed or pure for us) state $\rho$, the outcome $x$ \textit{will} be in the set $\{ x \; | \;  \hat{P}_{x}\rho \neq 0 \}$. 

We also assume a set of rationality constraints. First, we assume that preferences are transitive: where $X \succ Y$ means one prefers $X$ to $Y$, if $A \succ B$ and $B \succ C$, then $A \succ C$. This means we can order our preferences. Second, we assume that preferences can be assigned utility values $v$. Third, we assume additivity: an agent is indifferent between receiving two separate payoffs with utilities $x_1$ and $x_2$ and a single payoff $x_1 + x_2$. Fourth, we assume what Deutsch calls the two-player zero-sum rule, i.e. a game where payoffs consist of money changing hands between two players. That is, if A receives $x$ in one role, say, as `player' of the game, they receive $-x$ in the other role as `banker' instead. Finally, we assume substitutibility: for composite games, with sub-games, if any of the sub-games are replaced by a game of equal value, then to a rational player the value of the composite game is unchanged. 

With the set-up, we can consider generalized quantum games on states $\rho$ which may be pure or mixed. Given an observable $\hat{X}$ of the system in state $\rho$ which is to be measured, a player of such a game receives some payoff depending only on the outcome of measurement. For simplicity, we assume that the payoff utilities are numerically equivalent to the eigenvalues associated with the measurement outcomes, as Deutsch does: e.g. if the outcome is $|x_a\rangle\langle x_a|$ then the payoff utility is $x_a$. The valuation function for some given state $\rho$, $V[\rho]$, is the utility or value of playing a game with system in state $\rho$. Finally, by appealing to \textbf{Ignore Cross Terms} discussed in \S2.2, we make explicit what Deutsch employs only implicitly: the measurement outcomes correspond only to diagonal terms of a density matrix (pure or mixed); measurement outcomes never correspond to cross terms. With this set-up, Deutsch proves, assuming WFR\textsubscript{E}, that: 
\begin{equation}
 V[\rho] = \text{Tr}(\rho \hat{X}) = \sum_a x_a | \langle x_a | \psi \rangle |^2
\end{equation}
That is, the rational agent ought to bet on $\rho$ by weighing its expected value in accordance with the Born rule.

\subsubsection{Generalizing the Deutsch Proof to DMR\textsubscript{E}}

We now extend Deutsch's proof to DMR\textsubscript{E}. First we prove the Born rule for eigenstates of the observable $\hat{X}$ to be measured. Next, we consider equal superpositions of two eigenstates of $\hat{X}$. Then, we generalize to equal superpositions of arbitrary $N$ eigenstates. After that, we generalize to unequal superpositions of two eigenstates. Finally, we generalize to arbitrary real-valued coefficients in the superposition, completing the extension. 

\subsubsection*{Step 1: single eigenstate} 

This step is straightforward. Since $|\psi\rangle$ is, by hypothesis, in any eigenstate $|x_a\rangle$ of $\hat{X}$, no probabilities are required given our assumption of what `must happen' in quantum mechanics; the outcome of a measurement made in terms of $\hat{X}$ must have eigenvalue $x_a$, and therefore utility $x_a$, for a player playing a quantum game on such a state. Hence
\begin{equation}
    V \left[ |x_a\rangle\langle x_a| \right] = x_a
\end{equation}
That is, given $|x_a\rangle\langle x_a|$, the agent should expect outcome $x_a$ to occur with certainty, in line with the Born rule.

\subsubsection*{Step 2: equal weight superposition of two eigenstates of $\hat{X}$}

Now we prove the Born rule for equal superpositions of two eigenstates $|x_1 \rangle$ and $|x_2 \rangle$:
\begin{equation}\label{equal2superposition}
    \rho = \frac{1}{2} \bigg( |x_1 \rangle \langle x_1 | + |x_1 \rangle \langle x_2 | + |x_2 \rangle \langle x_1 | + |x_2 \rangle \langle x_2 |  \bigg)
\end{equation}
Note that the c.t., $|x_1 \rangle \langle x_2 | $ and $|x_2 \rangle \langle x_1 | $, show up again. However, they do not correspond to measurement outcomes. Our agent follows the \textbf{Ignore Cross Terms} rule. While this is explicit here, it is \textit{also} implicitly crucial to Deutsch's proof (since Deutsch \textit{also} assumes the obtaining of determinate measurement outcomes, to which DH is relevant). 

For such cases the goal is to prove
\begin{equation}
    V[\rho] = \frac{1}{2}\bigg(x_1 + x_2\bigg)
\end{equation}
hence showing that the agent ought to bet on each outcome as though they occur with probability 1/2, per the Born rule. 

Similar to Deutsch, we appeal to a consequence of additivity, suitably generalized to DMR\textsubscript{E}:
\begin{equation}\label{consequenceofadditivity}
    V\left[ \sum_a \sum_b \lambda_{ab} |x_a + k \rangle \langle x_b + k|\right] = k + V\left[ \sum_a \sum_b  \lambda_{ab} |x_a \rangle \langle x_b | \right]
\end{equation}
where $\lambda_{ab} = \psi_a \psi^*_b$. Recall that additivity demands indifference between receiving two separate payoffs with utilities $x_1$ and $x_2$ and a single payoff $x_1 + x_2$. Using this principle, we prove \eqref{consequenceofadditivity} in the same way as Deutsch, when we understand these payoffs in terms of possible observations given by non-cross terms. First, receiving any of the possible payoffs $x_a + k$ is the same as receiving $x_a$ first, and then $k$. This is the case even with c.t., which the agent neglects because they \textit{never} correspond to possible measurement outcomes.  Second, that's the same as playing a game with the state $\rho$, and \textit{then} just receiving a payoff $k$ directly. 

Now we use the zero-sum rule. The value of the game, of acting instead as banker in such a game (receiving $-x_a$ when measurement outcome is $x_a$), is the negative of the value of the original game. Hence:
\begin{equation}
    V\left[ \sum_a \sum_b  \lambda_{ab} |x_a \rangle \langle x_b | \right] + V\left[ \sum_a \sum_b  \lambda_{-a-b} |-x_a \rangle \langle -x_b | \right] = 0
\end{equation}
Given \eqref{equal2superposition}, it follows that:
\begin{equation}
\begin{aligned}
    & V\left[ \frac{1}{2} \bigg( |x_1 \rangle \langle x_1 | + |x_1 \rangle \langle x_2 | + |x_2 \rangle \langle x_1 | + |x_2 \rangle \langle x_2 |  \bigg) \right] \\  = - & V\left[\frac{1}{2} \bigg( | -x_1 \rangle \langle -x_1 | + |-x_1 \rangle \langle -x_2 | + |-x_2 \rangle \langle -x_1 | + |-x_2 \rangle \langle -x_2 |  \bigg) \right] 
\end{aligned}
\end{equation}
Using \eqref{consequenceofadditivity}, rewriting the right hand side using $k = -x_1 -x_2$, we get:
\begin{equation}
\begin{aligned}
        & V\left[ \frac{1}{2} \bigg( |x_1 \rangle \langle x_1 | + |x_1 \rangle \langle x_2 | + |x_2 \rangle \langle x_1 | + |x_2 \rangle \langle x_2 |  \bigg) \right] \\  = - &V\left[\frac{1}{2} \bigg( |x_2 + k \rangle \langle x_2+ k | + |x_2 + k \rangle \langle x_1 + k | + |x_1 + k \rangle \langle x_2 + k | + |x_1 + k \rangle \langle x_1 + k | \bigg) \right] 
\end{aligned}
\end{equation}
This entails:
\begin{equation}
\begin{aligned}
      & V\left[ \frac{1}{2} \bigg( |x_1 \rangle \langle x_1 | + |x_1 \rangle \langle x_2 | + |x_2 \rangle \langle x_1 | + |x_2 \rangle \langle x_2 |  \bigg) \right] \\ = - & k - V\left[\frac{1}{2} \bigg( | x_1 \rangle \langle x_1 | + |x_1 \rangle \langle x_2 | + |x_2 \rangle \langle x_1 | + |x_2 \rangle \langle x_2 |  \bigg) \right] 
\end{aligned}
\end{equation}
And since $k = -x_1 -x_2$:
\begin{equation}\label{equalsuperposition}
     V\left[ \frac{1}{2} \bigg( |x_1 \rangle \langle x_1 | + |x_1 \rangle \langle x_2 | + |x_2 \rangle \langle x_1 | + |x_2 \rangle \langle x_2 |  \bigg) \right] = \frac{1}{2}\bigg(x_1 + x_2\bigg)
\end{equation}
We arrive at what we sought to prove: the Born rule. $\square$ 

\subsubsection*{Step 3: equal weight superposition of $N$ eigenstates}

Now we want to generalize \eqref{equalsuperposition} and show:
\begin{equation}\label{generalequalsuperposition}
    V\left[ \frac{1}{N} \bigg( |x_1 \rangle \langle x_1 | + |x_2 \rangle \langle x_2 | + ... |x_N \rangle \langle x_N | + c.t. \bigg) \right] = \frac{1}{N}\bigg(x_1 + x_2 + ... + x_N\bigg)
\end{equation}
Generalizing Deutsch's procedure, we prove that this holds with $N = 2^m$, an even number for some non-zero integer $m$, by appealing to substitutibility. We start with the base case, where there is a game with just two equal weight outcomes (i.e. the game from the previous step). Then for each outcome, we simply replace it with a new sub-game with $2^{m-1}$ equal-weight outcomes, since agents will be indifferent between this new game and the original two-outcome game so long as the total value of the two games are identical. This covers the case where $N = 2^m$ for any $m$. Nothing turns on DMR\textsubscript{E} here. 

With the $N = 2^m$ cases covered, we now use this to generalize to the case of arbitrary integers $N$. Following Deutsch, observe that a game has value $v$ if every possible outcome of a game has a payoff $v$. So if $V\left[|\psi_1\rangle\langle\psi_1|\right] = V\left[|\psi_2\rangle\langle\psi_2|\right] = v$ (where $|\psi_1\rangle\langle\psi_1|$ and $|\psi_2\rangle\langle\psi_2|$ are superpositions of eigenstates of $\hat{X}$ chosen from two non-intersecting sets of eigenstates), then (by substitutability) the value of a game played on any convex combination of the two states, where $\alpha, \beta$ are complex numbers, also has value $v$:
\begin{equation}\label{convex}
    V\left[\frac{|\alpha|^2 |\psi_1\rangle\langle\psi_1| + |\beta|^2 |\psi_2\rangle\langle\psi_2|}{|\alpha|^2 + |\beta|^2 }  \right] = v 
\end{equation}
Then we set: 
\begin{equation}
    |\psi_1\rangle\langle\psi_1| = \frac{1}{N-1} \left( |x_1\rangle\langle x_1| + |x_2\rangle\langle x_2| + ... + |x_{N-1}\rangle\langle x_{N-1}| + c.t. \right)
\end{equation}
\begin{equation}
    |\psi_2\rangle\langle\psi_2| = |V\left[ |\psi_1\rangle\langle \psi_1 | \right] \rangle \langle V\left[ |\psi_1\rangle\langle \psi_1 | \right] | 
\end{equation}
That is, the eigenvalue associated with $|\psi_2\rangle\langle\psi_2|$ is the value of $  |\psi_1\rangle\langle\psi_1| $, and we also set:
\begin{equation*}
    \alpha = \sqrt{N - 1}
\end{equation*}
\begin{equation*}
    \beta = 1
\end{equation*}
Substituting these into \eqref{convex} entails:
\begin{equation}
\begin{aligned}\label{consequenceofconvex}
        & V\left[\frac{1}{N} \bigg( |x_1\rangle\langle x_1| + |x_2\rangle\langle x_2| + ... + |x_{N-1}\rangle\langle x_{N-1}| + |V\left[ |\psi_1\rangle\langle \psi_1 | \right] \rangle \langle V\left[ |\psi_1\rangle\langle \psi_1 | \right] | + c.t. \bigg)  \right] \\
        & = V\left[ |\psi_1\rangle\langle \psi_1 | \right]
\end{aligned}
\end{equation}
Note that this substitution is valid only if $V\left[ |\psi_1\rangle\langle \psi_1 | \right]$ is a different eigenvalue from other eigenvalues $x_1, ... , x_{N-1}$ (\citet[3134]{deutsch1999}).

Now we proceed inductively for our proof. Start by checking the base case, that \eqref{generalequalsuperposition} holds for some arbitrary $N = 2^m$. We've already proven that this is true above. Then we prove the inductive step, that if \eqref{generalequalsuperposition} holds for some $N = n$, then it holds for some $N = n - 1$. To do so, assume for the $n$\textsuperscript{th} case that the Born rule holds:
\begin{equation}\label{inductiveif}
\begin{aligned}
        & V\left[ \frac{1}{n} \bigg( |x_1 \rangle \langle x_1 | + |x_2 \rangle \langle x_2 | + ... + |x_{n-1} \rangle \langle x_{n-1} | + |x_{n} \rangle \langle x_{n} | + c.t. \bigg) \right] \\ & = \frac{1}{n}\bigg(x_1 + x_2 + ... + x_{n-1} + x_{n}\bigg) 
\end{aligned}
\end{equation}
Since it is different from other eigenvalues and eigenstates, we can rewrite $|V\left[ |\psi_1\rangle\langle \psi_1 | \right] \rangle \langle V\left[ |\psi_1\rangle\langle \psi_1 | \right] |$ as $|x_n\rangle \langle x_n|$ by relabelling `$V\left[ |\psi_1\rangle\langle \psi_1 | \right]$' as `$x_n$'. Then we see from \eqref{consequenceofconvex} and \eqref{inductiveif} that:
\begin{equation}
   \frac{1}{n}\bigg(x_1 + x_2 + ... + x_{n-1} + x_{n}\bigg) = V\left[ |\psi_1\rangle\langle \psi_1 |\right]
\end{equation}
or with a change of labels again:
\begin{equation}
    \frac{1}{n}\bigg(x_1 + x_2 + ... + x_{n-1} + V\left[ |\psi_1\rangle\langle \psi_1 | \right] \bigg) = V\left[ |\psi_1\rangle\langle \psi_1 |\right] 
\end{equation}
Multiplying by $n$, subtracting both sides by $ V\left[ |\psi_1\rangle\langle \psi_1 |\right]$, and dividing across by $n-1$, we get:
\begin{equation}
    V\left[ |\psi_1\rangle\langle \psi_1 | \right] = \frac{1}{n-1}\bigg(x_1 + x_2 + ... + x_{n-1}\bigg) 
\end{equation}
which is just \eqref{generalequalsuperposition} for $N = n-1$ given our definition of $|\psi_1\rangle\langle \psi_1 |$. This completes our proof for the inductive step. Since \eqref{generalequalsuperposition} holds for any arbitrary $N = 2^m$, and if it holds for some $N = n$ then it holds for $N = n-1$, it follows that \eqref{generalequalsuperposition} holds for all integers $N$. $\square$

\subsubsection*{Step 4: unequal weight superpositions of two eigenstates}

Now we consider systems in the unequal superposition of two eigenstates for all integers $N$, to prove the Born rule with arbitrary rational weights:
\begin{equation}
    \rho = \frac{1}{N} \bigg( j|x_1 \rangle \langle x_1 | + k |x_2 \rangle \langle x_2 | + c.t. \bigg) 
\end{equation}
We first prove the Born rule for a special case before specifying the general strategy. Consider the case of $N = 3, j = 1, k = 2$:
\begin{equation}
   \rho = \frac{1}{3} \bigg( |x_1 \rangle \langle x_1 | + 2 \, |x_2 \rangle \langle x_2 | + c.t. \bigg) 
\end{equation}
In this case, we want to prove that:
\begin{equation}
    V\left[   \frac{1}{3}|x_1 \rangle \langle x_1 | +  \frac{2}{3}|x_2 \rangle \langle x_2 | + c.t.  \right] = \frac{x_1 + 2x_2}{3}
\end{equation}
To do so, consider auxiliary devices $Y$ in state $\rho_Y$ depending on the eigenvalues obtained when a measurement on $\hat{M}$ is made on the system $S$:
\begin{equation}
\hat{M}=
\begin{cases}
x_1, \; \, \rho_Y = |Y_1\rangle \langle Y_1| \\
x_2, \; \; \rho_Y = \frac{1}{2} \bigg( |Y_2\rangle \langle Y_2| + |Y_3\rangle \langle Y_3| + c.t. \bigg)
\end{cases}
\end{equation}
such that the joint state of the universe -- including $S$ and $Y$ -- after measurement is: 
\begin{equation}\label{equalweightn}
    \rho_{S + Y} = \frac{1}{3} \bigg( |x_1\rangle|Y_1\rangle \langle Y_1|\langle x_1| + |x_2\rangle|Y_2\rangle \langle Y_2|\langle x_2| + |x_2\rangle|Y_3\rangle \langle Y_3|\langle x_2| + c.t. \bigg)
\end{equation}
Then we make a measurement $\hat{Y}$ for the auxiliary system: if we obtain $Y_1$, then $\hat{M}$ has value $x_1$. Otherwise, $\hat{M}$ has value $A$. Following Deutsch, we can also demand that the sum of $\hat{Y}$'s eigenvalues vanish, which amounts to stating that players will be indifferent to playing a further game in which they receive the measured value of $\hat{Y}$. That is, the outcome of measuring $\hat{Y}$ doesn't matter to them:
\begin{equation}
     Y_1 = Y_2 + Y_3 = 0
\end{equation}
Hence, a game played on the same state using observables $\hat{M}$ and $\hat{Y}$ has the same value as one using just $\hat{M}$. Because of additivity, the composite game also has the same value as a game where only one measurement is made in terms of the observable $\hat{M} \otimes \hat{1} + \hat{1} \otimes \hat{Y}$. In terms of that observable, $\rho_{S+Y}$ is just:
\begin{equation}
    \frac{1}{3} \bigg( |x_1 + Y_1\rangle \langle Y_1 + x_1| + |x_2 + Y_2\rangle \langle Y_2 + x_2| + |x_2 + Y_3\rangle \langle Y_3 + x_2| + c.t. \bigg)
\end{equation}
which is an equal superposition of three eigenstates, allowing us to use the results of step 3. Together with the constraint on $\hat{Y}$'s eigenvalues vanishing, we get: 
\begin{equation}
\begin{aligned}
    & V \left[ \frac{1}{3} \bigg( |x_1 + Y_1\rangle \langle Y_1 + x_1| + |x_2 + Y_2\rangle \langle Y_2 + x_2| + |x_2 + Y_3\rangle \langle Y_3 + x_2| + c.t. \bigg) \right ] \\ & = \frac{1}{3} \bigg( (x_1 + Y_1) + (x_2 + Y_2) + (x_2 + Y_3) \bigg) = \frac{x_1 + 2x_2}{3}
\end{aligned}
\end{equation}
which gives us exactly what we were looking for. To generalize to weights of arbitrary rational numbers:
\begin{equation}
    V\left[   \frac{j}{N}|x_1 \rangle \langle x_1 | +  \frac{k}{N}|x_2 \rangle \langle x_2 | + c.t. \right] = \frac{jx_1 + kx_2}{N}
\end{equation}
we can simply generalize our choice of auxiliary devices such that: 
\begin{equation}
\hat{X}=
\begin{cases}
x_1, \rho_Y = \frac{1}{j} \sum^j_{a = 1} |Y_a\rangle \langle Y_a| \\
x_2, \rho_Y = \frac{1}{k} \sum^N_{a = j + 1} |Y_a\rangle \langle Y_a|
\end{cases}
\end{equation}
to arrive at a generalized version of \eqref{equalweightn}. The same procedure can then be applied. $\square$ 


\x{\subsubsection{Putting them together}}
Finally, we prove the Born rule for arbitrary real weights. One strategy, due to \citet[68]{sebenscarroll2016}, is to observe that, for any wavefunction with irrational weights, there exists arbitrarily similar wavefunctions with rational weights since the rationals are dense in the reals assuming the continuous variation of probabilities given small changes in weights.

A related strategy, in line with the decision-theoretic program, is due to \citet[3135--3136]{deutsch1999}. Consider games where agents are given systems in state $\rho$ (with irrational weights) to bet on, but before measurement on $\hat{X}$ the system undergoes unitary evolution $U$ such that the new state is $U \rho U^\dagger$. If each eigenstate of $\rho$ thereby evolves unitarily to a superposition of higher (or lower) eigenvalue eigenstates, the total value of the game on $U \rho U^\dagger$ exceeds (or is exceeded by) the value of a game on $\rho$, by additivity and substitutibility. We can use such transformations to figure out the upper and lower bounds (in rational numbers) of the values of games on $\rho$. 

Appealing to the denseness of the rationals in the reals, there exists arbitrarily slight unitary transformations which act on $\rho$ in two ways: (i) transformations $U_{\uparrow}$ which take each eigenstate (in terms of $\hat{X}$) in the expansion of $\rho$ to superpositions of itself and \textit{higher}-eigenvalue eigenstates in $U_{\uparrow} \rho U_{\uparrow}^\dagger$ with rational weights, and (ii) transformations $U_{\downarrow}$ which take each eigenstate in $\rho$ to superpositions of itself and \textit{lower}-eigenvalue eigenstates in $U_{\downarrow} \rho U_{\downarrow}^\dagger$ with rational weights. 

Games played with $U_{\uparrow} \rho U_{\uparrow}^\dagger$ are at \textit{least} as valuable as the original game such that $ V \left[U_{\uparrow} \rho U_{\uparrow}^\dagger\right] \succeq V \left[ \rho \right]$, and the rational values of the former can be made arbitrarily close to $\text{Tr}(\rho \hat{X})$.   Similarly, games played with $U_{\downarrow} \rho U_{\downarrow}^\dagger$ are at \textit{most} as valuable as the original game such that $ V \left[U_{\downarrow} \rho U_{\downarrow}^\dagger \right] \preceq V \left[ \rho \right]$, , and the rational values of the former can be made arbitrarily close to $\text{Tr}(\rho \hat{X})$.   Since $ V \left[U_{\uparrow} \rho U_{\uparrow}^\dagger\right] \succeq V \left[ \rho \right] \succeq V \left[U_{\downarrow} \rho U_{\downarrow}^\dagger \right]$, and $V \left[U_{\uparrow} \rho U_{\uparrow}^\dagger\right]$ and $V \left[U_{\downarrow} \rho U_{\downarrow}^\dagger \right]$ both approach $\text{Tr}(\rho \hat{X})$ in the limit, it follows that
\begin{equation}
    V \left[ \rho \right] = \text{Tr}(\rho \hat{X})
\end{equation}
Even with irrational weights, and hence for arbitrary \textit{real} weights, an agent betting on $\rho$ should bet in accordance with the Born rule. $\square$

In contrast to Deutsch's proof, we skip the generalization to complex-valued amplitudes, because the coefficients of branches in the expansion of $\rho$ are always real-valued.

This completes our decision-theoretic proof of the Born rule for DMR\textsubscript{E}. Rational agents, knowing the state of the world $\rho$, the observables $\hat{X}$ to be used, equipped with the appropriate quantum-mechanical and rationality assumptions, should adopt the Born rule when it comes to playing quantum games in \textit{both} WFR\textsubscript{E} and DMR\textsubscript{E}. Fundamental insights in Deutsch's decision-theoretic proof, like those in the other programs, do not depend on $\rho$ being pure.

\section{Discussion}

We've generalized several arguments for branching and the Born rule from WFR\textsubscript{E} to DMR\textsubscript{E}.  In addition to answering open questions in the literature, we take our results to have further conceptual implications. 

First, in order to set up the generalized arguments, we needed to contemplate, without presupposing a universal pure state, the ontological structure of the Everettian multiverse. For EQM to allow both WFR and DMR, the story about decoherence and branching should apply to both without prejudice. As we've seen, that is indeed the case. This leads us to see that the essence of the Everettian story about the emergence of a multiverse is not a universal wavefunction that gives rise to many branches represented by wavefunctions, but a universal density matrix (which can be pure or mixed) that gives rise to many branches represented by density matrices.  According to the perspective of DMR\textsubscript{E}, a pure-state multiverse is a very special case. 

Second, with DMR\textsubscript{E}, Everettians can explore new theoretical possibilities of DMR. \x{This can be helpful in quantum cosmology, where several proposals require a fundamental mixed state (\citet{page1986density, page2008no, barvinsky2006cosmological}), which is allowed by DMR but disallowed by WFR.} On a more philosophical level, we can consider a unified treatment of `classical' and `quantum' probabilities in EQM. In WFR\textsubscript{E}, there are two fundamental sources of probabilities: the quantum probability of finding ourselves in a particular branch \y{(or betting preferences in the decision-theoretic framework)}, associated with the weight of the branch in the multiverse, and the classical probability of the particular multiverse, associated with a density matrix representing our ignorance of the underlying universal pure state. Their justifications are very different. \y{The latter is not understood as standard quantum probabilities, justifiable in terms of self-locating uncertainties or betting preferences. Instead, it may have a statistical-mechanical origin, corresponding to a probability distribution over initial universal quantum states, the so-called \textit{Statistical Postulate} (\cite{albert2000time}).}   In DMR\textsubscript{E}, however, the two \textit{can} be reduced to a single notion of probability, that of finding ourselves in a particular branch, albeit in a more expansive multiverse. Whichever $\rho$ is used by defenders of WFR\textsubscript{E} to represent their ignorance of the fundamental pure state of the multiverse, defenders of DMR\textsubscript{E} can regard that $\rho$ as the fundamental mixed state. Insofar as classical and quantum probabilities in EQM can be reduced to a single source, they also can be justified in the same way.

A theory on which we can apply this strategy is the Everettian Wentaculus (\cite{chen2018, chen2022strong, ChenNature2023}). This version of DMR\textsubscript{E} proposes a simple and unique choice of the initial density matrix of the multiverse (as a version of the Past Hypothesis) and regards it as the only nomological possibility. As a matter of physical laws, the history of the Everettian multiverse could not have been different. There is no longer a choice of the fundamental density matrix, beyond the choice of the physical law, because the actual one is nomologically necessary. It is an instance of ``strong determinism.'' Both classical (statistical-mechanical) and quantum probabilities can be understood as branch weights of the Everettian Wentaculus multiverse, represented by a mixed-state density matrix. \y{With the possibility of a unified treatment of probabilities (\x{in the sense described earlier}), the generalization from WFR\textsubscript{E} to DMR\textsubscript{E} is theoretically attractive.}\footnote{For two other proposals of eliminating the Statistical Postulate, see Albert (\citeyear{albert2000time}, \S7) and Wallace (2012, \S9).}

Finally, in both the self-location framework and Deutsch's version of the decision-theoretic proof, we've derived the Born rule in DMR\textsubscript{E} in ways similar to WFR\textsubscript{E}, by appealing to the same principles (separability and symmetry on one hand, decision-theoretic principles on the other) and metaphysical foundations (decoherence and branching). The two theories are empirically equivalent, not just mathematically, but also conceptually. They give us the same empirical predictions and the same kind of probabilities (based on self-locating uncertainties or decision-theoretic preferences). Everettians, by their own lights, may regard DMR\textsubscript{E} as a genuine alternative to WFR\textsubscript{E}. Everettians who are attracted to those frameworks, then, face the question which version of EQM they should accept. The answer cannot be based on experimental results or the lack of solutions to conceptual problems of DMR\textsubscript{E}.   

\xx{We emphasize that these considerations do not resolve debates about ontology and probability in EQM. To the extent that the proposed solutions are effective within  WFR\textsubscript{E}, they can be extended to DMR\textsubscript{E}. However, the same foundational issues remain open to discussion within this more general framework.} 

\section{Conclusion}

We suggest that several Everettian justifications for decoherence, branching, and the Born rule apply to both WFR\textsubscript{E} and DMR\textsubscript{E}. Hence, the theoretical benefits of DMR are available to EQM. Everettians face a choice between two types of theories, one allowing only pure states for the multiverse and the other also allowing mixed states. The availability of different versions of EQM is an interesting example of empirical underdetermination. Its implications and possible resolutions are questions we leave for future work. 

\section*{Acknowledgements}
For helpful discussions, we thank Jeffrey Barrett, Kelvin McQueen, Katie Robertson, Simon Saunders, Charles Sebens, Tony Short, Karim Th\'ebault, and David Wallace, several anonymous referees, and the participants at the Philosophy of Physics Society meeting at the 2024 Pacific APA, the 2023 Workshop on Relational Clocks, Decoherence, and the Arrow of Time at the University of Bristol, and the 2022 California Quantum Interpretation Network Conference at Chapman University.

\bibliographystyle{plainnat}

\bibliography{bib.bib}

\end{document}